\documentclass{article}
\usepackage{amsmath}
\usepackage{amssymb}
\usepackage{dsfont}
\usepackage{graphicx}
\usepackage{xcolor}

\title{Pre- and post-processing in quantum-computational hydrologic inverse analysis}
\author{John K. Golden$^{1,*}$, Daniel O'Malley$^{1,2}$\\
{\footnotesize$^1$Computational Earth Science, Los Alamos National Laboratory, Los Alamos, NM 87545}\\
{\footnotesize$^2$Department of Computer Science and Electrical Engineering, University of Maryland,}\\{\footnotesize Baltimore County, MD 21250}\\{\footnotesize$^*$ Corresponding author: golden@lanl.gov}}

\begin{document}

\maketitle

\begin{abstract}
It was recently shown that certain subsurface hydrological inverse problems -- here framed as determining the composition of an aquifer from pressure readings -- can be solved on a quantum annealer. 
However, the quantum annealer performance suffered when solving problems where the aquifer was composed of materials with vastly different permeability, which is often encountered in practice.
In this paper, we study why this regime is difficult and use several pre- and post-processing tools to address these issues.  
This study has three benefits: it improves quantum annealing performance for real-world problems in hydrology, it studies the scaling behavior for these problems (which were previously studied at a fixed size), and it elucidates a challenging class of problems that are amenable to quantum annealers.

\end{abstract}

\section{Introduction}

The recent introduction of quantum annealer hardware with thousands of qubits has opened a door to solving discrete optimization problems in new ways.
Hardware such as the D-Wave 2000Q aims to solve quadratic unconstrained binary optimization problems (QUBOs) containing up to 2048 variables and have sparse quadratic terms.
Smaller problems that are less sparse can be solved by using several qubits to represent one logical variable in the discrete optimization problem.
It was previously shown that certain subsurface hydrologic inverse problems \cite{omalley2018hydro} fit naturally in this problem formulation.
Solving these hydrologic inverse problems is critical since the parameters (spatially heterogeneous permeabilities) that determine how fluid flows in the earth's subsurface cannot be readily observed and must be determined through an inverse analysis.
Understanding these flows is essential to applications such as geologic carbon sequestration \cite{class2009benchmark,harp2017development}, groundwater contaminant remediation \cite{mackay1989groundwater,o2014combined}, and energy production \cite{o2016does,hyman2016understanding}.
Methods of formulating and solving these inverse problems remains an active area of research \cite{lee2014large,lin2016computationally,lin2017large,mo2019deep,o2019learning}, and quantum annealing provides an interesting possibility that impacts both the formulation of the problem and method of solution.

A conceptual formulation of the hydraulic inverse problem that is well-suited to quantum annealing involves decomposing an aquifer into two regions, one with high permeability ($k_H$) and one with low permeability ($k_L$).
In \cite{omalley2018hydro} it was observed that the performance of the quantum annealer degraded as the difference in permeability between the two materials, $\Delta k=k_H-k_L$, increased.
This observation is important for two reasons. 
On the hydrological side, it is important because real-world aquifers are often composed of materials with vastly different permeability, e.g. clay vs. sand. 
Therefore using a quantum annealer to solve realistic problems requires an improvement in performance for large $\Delta k$.
And on the quantum annealing and mathematical side, it is interesting to study why large $\Delta k$ is more difficult than small $\Delta k$.
In particular, physical intuition would suggest that it is easier to observe differences in an aquifer composed of very different materials.
Hence, it is surprising that the quantum annealer has increased difficulty in this regime.

The primary purpose of this paper is, therefore, to discuss the large $\Delta k$ behavior in greater detail, and use it as an opportunity to study several classical tools for improving results from the quantum annealer.
The three techniques we study have been selected for their general applicability, and to provide a cross-section of the different types of pre- and post-processing techniques available to researchers working with the quantum annealer hardware and any other future quantum annealers. 
For a pre-processing algorithm, we employ a long-standing technique based on the Roof-dual and strong persistence \cite{boros2008max}. 
This uses graph algorithms and algebraic manipulations to discover the ``low-hanging fruit'' of the optimization problem, i.e. variables which must take certain values in the optimal answer. 
We study two post-processing techniques. 
The first, post-process optimization, is built into the D-Wave application programming interface (API). It employs a fast, exact solver on local parts of the problem in conjunction with the quantum annealer hardware for global exploration.
The second, multi-qubit correction (MQC), was introduced in the literature recently \cite{dorband2018method}.
MQC takes a collection of samples from the quantum annealer, breaks them into sub-samples, and then amalgamates them into an improved single sample. 
More specifics of these techniques are in Section~\ref{sec:algorithms}.

In addition to studying the scaling behavior as $\Delta k$ becomes large, we also explore the scaling behavior as the size hydrological inverse problem grows (here measured as an $N\times N$ grid of groundwater pressure measurements).
The scaling in $N$ gives some indication of the long-term potential of quantum annealing for solving these problems, as large grids are often used in practice.
However, we consider the scaling in $N$ to be of secondary importance to the scaling in $\Delta k$.
This is because advancing the methods so that they can handle realistically large $\Delta k$ values on small grids is needed before focusing on very large grids (for which sufficiently large quantum annealers also do not currently exist).

Also discussed in \cite{omalley2018hydro} was the effect of observational noise (i.e., noise in the hydrologic observations) on the performance of the quantum annealer. 
In this paper, we also analyze more thoroughly how noise impacts the problem as posed to the quantum annealer and the effect on the pre- and post-processing algorithms. 

This paper is structured as follows. Section~\ref{sec:why-large-k-hard} reviews the QUBO formulation of the hydrological inverse problem and then analytically shows why the large $\Delta k$ limit is more difficult to solve. We then introduce the pre- and post-processing techniques of interest in Section~\ref{sec:algorithms}, and show how they perform and scale in Section~\ref{sec:results}. The effects of observational noise are explored in Section~\ref{sec:noise}, and conclusions are detailed in Section~\ref{sec:conclusion}.

\section{Why is large $\Delta k$ hard?}\label{sec:why-large-k-hard}

Using $\mathbf{k}$ and $\mathbf{h}$ to denote the vectors composed by the permeability and hydraulic head (i.e. pressure readings from wells spread across the aquifer), the relationship between them is then expressed through a partial differential equation (PDE):
\begin{equation}\label{eq:pressure}
    \nabla \cdot (\mathbf{k} \nabla \mathbf{h}) = 0.
\end{equation}
This PDE can be discretized in numerous ways, and we exploit a finite difference approximation, as is done in previous work \cite{omalley2018hydro}.
In 1D, this discretization is
\begin{equation}
    \frac{-k_{i-1}h_{i-1} + 2(k_{i}+k_{i-1})h_i - k_ih_{i + 1}}{(\Delta x)^2} = 0
\end{equation}
where $h_i$ denotes the pressure at node $i$, $k_i$ denotes the permeability between nodes $i$ and $i + 1$, and $\Delta x$ is the distance between nodes.
Similarly, in 2D, this discretization is
\begin{multline}
    \frac{-k^x_{i-1,j}h_{i-1,j} + 2(k^x_{i,j}+k^x_{i-1,j})h_{i,j} - k^x_{i,j}h_{i + 1,j}}{(\Delta x)^2} \\+ \frac{-k^y_{i,j-1}h_{i,j-1} + 2(k^y_{i,j}+k^y_{i,j-1})h_{i,j} - k^y_{i,j}h_{i,j + 1}}{(\Delta y)^2} = 0
\end{multline}
where the pressures, $h_{i,j}$ now have two subscripts to denote the row and column of the node on a 2D grid.
Note that the 2D discretization involves an anisotropic permeability with $k^x_{i,j}$ denoting the permeability between pressure nodes $h_{i,j}$ and $h_{i+1,j}$ and $k^y_{i,j}$ denoting the permeability between pressure nodes $h_{i,j}$ and $h_{i,j+1}$.
In the context of the 2D problem, we use the notation $\mathbf{k}$ to denote a vectorization of these two 2D fields.

We generate synthetic instances of inverse problems by first sampling a vector of permeabilities, $\mathbf{k}_{true}$.
The components of $\mathbf{k}_{true}$ are treated as independent, identically distributed random variables with a $1/2$ probability of being either $k_L$ or $k_H$.
This is therefore an instance of a binary hydrological inverse problem, with only two possible permeabilities at each location, but such a model is, in fact, relevant for many real-world scenarios. 
The permeabilities are then used to solve the discretized form of eq.~(\ref{eq:pressure}) to obtain a set of hydraulic head observations.
These observations are then used to inform the inverse analysis.
In some cases, noise is added to the hydraulic head observations before performing the inverse analysis, and further details on the noise are provided in Section \ref{sec:noise}.
 
It was shown in \cite{omalley2018hydro} that given some $\mathbf{h}$, the $\mathbf{k}$ that solves eq.~(\ref{eq:pressure}) will also minimize the function
\begin{equation}\label{eq:dwave-hamiltonian}
    f(\mathbf{q}) = \sum_{i=1}^n a_i q_i + \sum_{i=1}^{n-1}b_i q_i q_{i+1}
\end{equation}
where 
\begin{eqnarray}\label{eq:qubo-coeffs}
    q_i &=& \frac{(k_i - k_L)}{(k_H - k_L)},\\
    a_i &=& \Delta h_i \left(-k_L (\Delta h_{i-1} - 2\Delta h_i +  \Delta h_{i+1}) + \Delta k \Delta h_i\right),\label{eq:qubo-coeffs-a} \\
    b_i &=& -\Delta h_i \Delta h_{i+1} \Delta k.\label{eq:qubo-coeffs-b}
\end{eqnarray}
The form of eq.~(\ref{eq:dwave-hamiltonian}) is that of a quadratic unconstrained optimization problem (QUBO), which the D-Wave quantum annealer is designed to solve.

\begin{figure}[h!]
    \includegraphics[width=\textwidth]{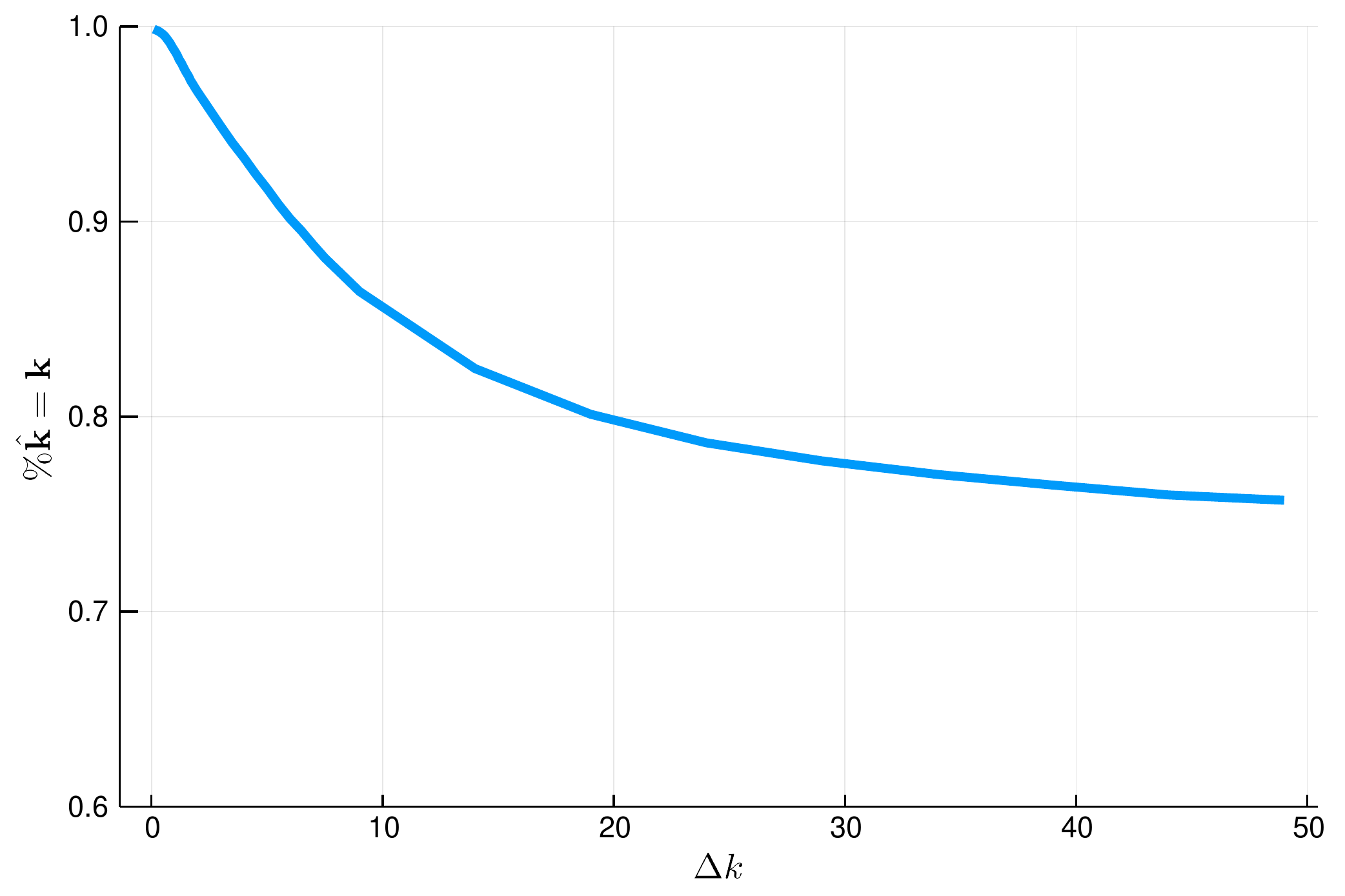}
    \caption{Accuracy of the quantum annealer for the 1D hydrological inverse problem with 2030 $\mathbf{k}$ values. The quantum annealer correctly determines a smaller percentage of the $\mathbf{k}$ as $\Delta k$ increases. This is because some terms in the QUBO are proportional to $(\Delta k)^{-1}$.}
    \label{fig:1d-delta-k}
\end{figure}

Now let us discuss the behavior of eq.~(\ref{eq:dwave-hamiltonian}) in the large $\Delta k$ limit. 
First, note that if one defines
\begin{equation}
    \Delta h_i = h_{i+1} - h_i
\end{equation}
then using eq.~(\ref{eq:pressure}) in the noise-free 1D binary context, it is easy to see that $\Delta h_i$ also takes binary values: $\Delta h_i \in \{\Delta h_L, \Delta h_H\}$, where
\begin{equation}\label{eq:delta-h-defs}
    \Delta h_H = -\frac{n k_L}{n_1 k_L + n_2 k_H}, ~~ \Delta h_L = -\frac{n k_H}{n_1 k_L + n_2 k_H}.
\end{equation}
Here, $n_1$ and $n_2$ represent the number of blocks of material with low and high permeability, respectively. 
Assuming that $n_1$ and $n_2$ are both $\mathcal{O}(n/2)$, then in the case $k_H \gg k_L$ (i.e., large $\Delta k$) we have
\begin{equation}
    \Delta h_H = \mathcal{O}(k_H^{-1}), ~~\Delta h_L = \mathcal{O}(1).
\end{equation}
Note that there are only 3 possible values $b_i$ can take:
\begin{equation}
    b_i \in \{ -\Delta h^2_H \Delta k, -\Delta h^2_L\Delta k, -\Delta h_H \Delta h_L\Delta k \}.
\end{equation}
In the case $k_H \gg k_L$, looking back at eq.~(\ref{eq:delta-h-defs}) we see that these translate to 
\begin{equation}\label{eq:coeff-kHigh-limit}
    b_i = \{ \mathcal{O}(k_H^{-1}), \mathcal{O}(k_H), \mathcal{O}(1) \}.
\end{equation}
We therefore see that terms in eq.~(\ref{eq:dwave-hamiltonian}) proportional to $\Delta h^2_H \Delta k$ will be orders of magnitude smaller than those proportional to $\Delta h^2_L \Delta k$.
While we do not enumerate all 8 possible values for $a_i$, we note that half of them are $\mathcal{O}(k_H)$ while the other half are $\mathcal{O}(k_H^{-1})$. 
Given the existence of these small coefficients, it is not surprising that a random realization of the QUBO in eq.~(\ref{eq:dwave-hamiltonian}) will feature some $q_i$ that are only multiplied by small $a_i$ and $b_i$ while other $q_i$ have much larger coefficients. 
From the perspective of the quantum annealer, the smaller terms are indistinguishable from hardware noise when $\Delta k$ is sufficiently large.
The quantum annealer therefore has little or no sensitivity to the $q_i$ with very small coefficients and effectively chooses their value at random, thus getting them right 50\% of the time. 
This behavior is captured in Figure~\ref{fig:1d-delta-k}.
Note that once $\Delta k$ gets past a certain point, the accuracy of the quantum annealer stops getting noticeably worse.
This is because once terms with small coefficients get so small as to become noise, it no longer matters how much smaller they get.

The main takeaway from this section is that large $\Delta k$ is difficult for the quantum annealer hardware because of the unique form of the QUBO formulation of the hydrological inverse problem.
As $\Delta k$ increases, the quantum annealer becomes more sensitive to some terms while increasingly ignoring others.

\section{Pre- and post-processing algorithms}\label{sec:algorithms}

In this section, we describe the three techniques that we apply to improve the performance of the quantum annealer for large $\Delta k$.
As mentioned in the introduction, these algorithms have been chosen for their widespread applicability.
We briefly discuss the design of a custom algorithm for this problem in Section~\ref{sec:results-2d}.

\subsection{Roof-duality \& strong persistence}

Roof-duality and strong persistence are components of a well-developed set of pre-processing techniques introduced in \cite{boros2006preprocessing}.
This algorithm runs in polynomial time, but with a fairly large exponent -- $O(n^4)$ where $n$ is the number of variables in the QUBO.
The goal of this analysis is to find the low-hanging fruit of the QUBO, that is, variables that these polynomial-time algorithms can determine must take certain values in any global minimum. 
These values are determined through a combination of derivative information as well as an implementation of max-flow. 
The same techniques can also give lower bounds on the minimum value of the QUBO, although we do not make use of that feature here. 

The D-Wave API has a built-in function that determines the values of these ``easy to fix'' variables, appropriately called \verb|fix_variables()|.
The variables whose values can be determined can then be removed from the QUBO, reducing its size and potentially improving the dynamic range of the QUBO that is ultimately submitted to the quantum annealer.
For the rest of this paper, we refer to this as the FV algorithm. 

\subsection{postprocess=``optimization''}

The \verb|postprocess="optimization"| (PO) option that we utilize is implemented in software that executes on classical hardware close to the quantum annealer (i.e., on the server).
PO works by decomposing the D-Wave Chimera graph into subgraphs, each with low treewidth.
An exact solver is then used to solve the localized version of the sub-QUBO associated with each of these graphs.
Each of these sub-QUBOs is localized at the point sampled from the annealer.
In the case of the ``plain'' results for the 2D problem, the default postprocessing is used for the virtual full yield solver, which is the \verb|postprocess="sampling"| option.
This option operates in a similar way to the PO option, with the first step being the decomposition into subgraphs with low treewidth.
However, instead of using an exact solver to solve the localized sub-QUBO, it uses a sampling algorithm to approximately sample from a Boltzmann distribution where the sub-QUBO is the energy.
Further details on these postprocessing options are available in D-Wave's postprocessing documentation \cite{dwave2018postprocessing}.

\subsection{Multi-Qubit Correction}

Multi-Qubit Correction (MQC) \cite{dorband2018method} is a postprocessing method that transforms a set of samples into a single sample that is at least as good (and often better) than the best sample.
The method works by first using two samples (call them $q^A$ and $q^B$) to construct a third sample (call it $q^C$) that is as good as or better than either of the original samples.
To understand the mechanics of the method, let $G$ be the graph associated with a QUBO with each qubit corresponding to a vertex and each nonzero quadratic term in the QUBO corresponding to an edge.
Now let $G'$ be the subgraph of $G$ containing the vertices that correspond to qubits where $q^A$ and $q^B$ differ (as well as all edges connecting these vertices).
The graph $G'$ is then dissected into its connected components $G'_1, G'_2,\ldots,G'_n$.
The constructed sample, $q^C$ is then defined by the following algorithm:
first, for each $i$ such that $q^A_i=q^B_i$ let $q^C_i=q^A_i=q^B_i$.
This defines $q^C$ on the qubits where $q^A$ and $q^B$ agree.
Second, for each connected component (call it $G'_i$) of $G'$ choose either $q^C_{G'_i}=q^A_{G'_i}$ or $q^C_{G'_i}=q^B_{G'_i}$ depending on which results in a lower QUBO value.
Note that each of these choices can be made independently because there are no couplers between the qubits in $G'_i$ and $G'_j$ when $i\ne j$.
Combining pairs of samples in this fashion reduces the number of samples by a factor of two.
This process is performed repeatedly, reducing the number of samples by a factor of two each time until only one sample remains.
Our implementation of the MQC algorithm has scaling $O(n^2 \log m)$ where $n$ is the number of qubits and $m$ is the number of samples.
Further details on the MQC algorithm are available \cite{dorband2018method}.

\section{Results}\label{sec:results}

In this section, we describe the results of using the FV, PO, and MQC algorithms to improve the quantum annealer's results. 
We then study how the speed and accuracy of the algorithms scale with problem size. 
Note that the results in this section do not incorporate any observational noise, i.e. the ground state of the QUBO exactly corresponds with the ground truth of the aquifer. 
In a later section, we discuss the impacts of observational noise on the methods presented here.
We also point out that we performed a study of the impact of varying the annealing time.
However, the annealing time had no discernable impact on the performance, so we omit those results except to mention them here.

\subsection{Methodology}\label{sec:methodology}

For this analysis, we employed the 1D and 2D embeddings and QUBO formulations introduced in~\cite{omalley2018hydro} -- the only difference being that we use the larger D-Wave 2000Q here as opposed to the D-Wave 2X that was used in the reference.
Our analysis was conducted on the D-Wave 2000Q machine at Los Alamos National Laboratories (LANL). 
This machine has 2032 active and 16 inactive qubits. 
The 1D problem is embedded on a simple linear chain $\{q_i\}$ of connected qubits, with each $q_i$ corresponding to a permeability.
The LANL D-Wave 2000Q has a chain of 2030 connected qubits so that we can embed a problem with up to 2030 permeability variables on the hardware QPU. 

The 2D problem requires a more complicated embedding to infer the anisotropic permeability in both $x$ and $y$ coordinate directions.
The embedding we employ utilizes a collection of 8 qubits (known as a ``unit cell'' in the Chimera topology of the D-Wave 2000Q) to encode the 4 permeabilities surrounding a single pressure node.
Each permeability is located between two pressure nodes, so it surrounds two pressure nodes (except for permeabilities on the boundary).
This means that each permeability variable that is not on the boundary is represented with 4 qubits.
To utilize the full $16\times16$ grid of unit cells in the D-Wave 2000Q chip, we use the Virtual Full-Yield Chimera (VFYC) solver~\cite{dwave2018postprocessing} to accommodate the missing qubits. 
Generally, a hydrologic inverse problem involving $N\times N$ unit cells uses all $8N^2$ qubits on an $N\times N$ Chimera graph and involves $2N(N+1)$ logical variables (i.e., permeabilities).

We only study 1D chains of maximal length and utilize the more difficult (and more realistic) 2D scenario to study how the pre- and post-processing algorithms scale with problem size. 
For both the 1D and 2D problems, our tests begin with synthetic data, in this case permeability values, $\mathbf{k}$, randomly chosen from $\{k_L=1, k_H\}$ with equal probability.
The hydraulic head measurements $\mathbf{h}$ are calculated from $\mathbf{k}$ and used to create the QUBO. 
We test the efficacy of the D-Wave (in combination with the pre- and post-processing algorithms under analysis) by determining what percentage of the $\mathbf{k}$ are correctly determined.
Unless otherwise noted (specifically in the case of the MQC algorithm), each D-Wave run generates 1000 possible solutions, from which we take the solution with the lowest energy in the corresponding QUBO.

\subsection{1D}

The results of the quantum annealer hardware without any pre- or post-processing for the 1D case are described in Figure~\ref{fig:1d-delta-k}. 

FV and PO are both able to find the correct values for all of the $\mathbf{k}$ for any 1D problem.
This is not surprising, as the 1D problem is, in fact, straightforward: a block with $k_i = k_H$ corresponds directly to $\Delta h_i = \Delta h_L$ and vice versa. 
In this sense, the 1D problem produces a QUBO that is readily handled by these techniques.

\begin{figure}[h!]
    \includegraphics[width=\textwidth]{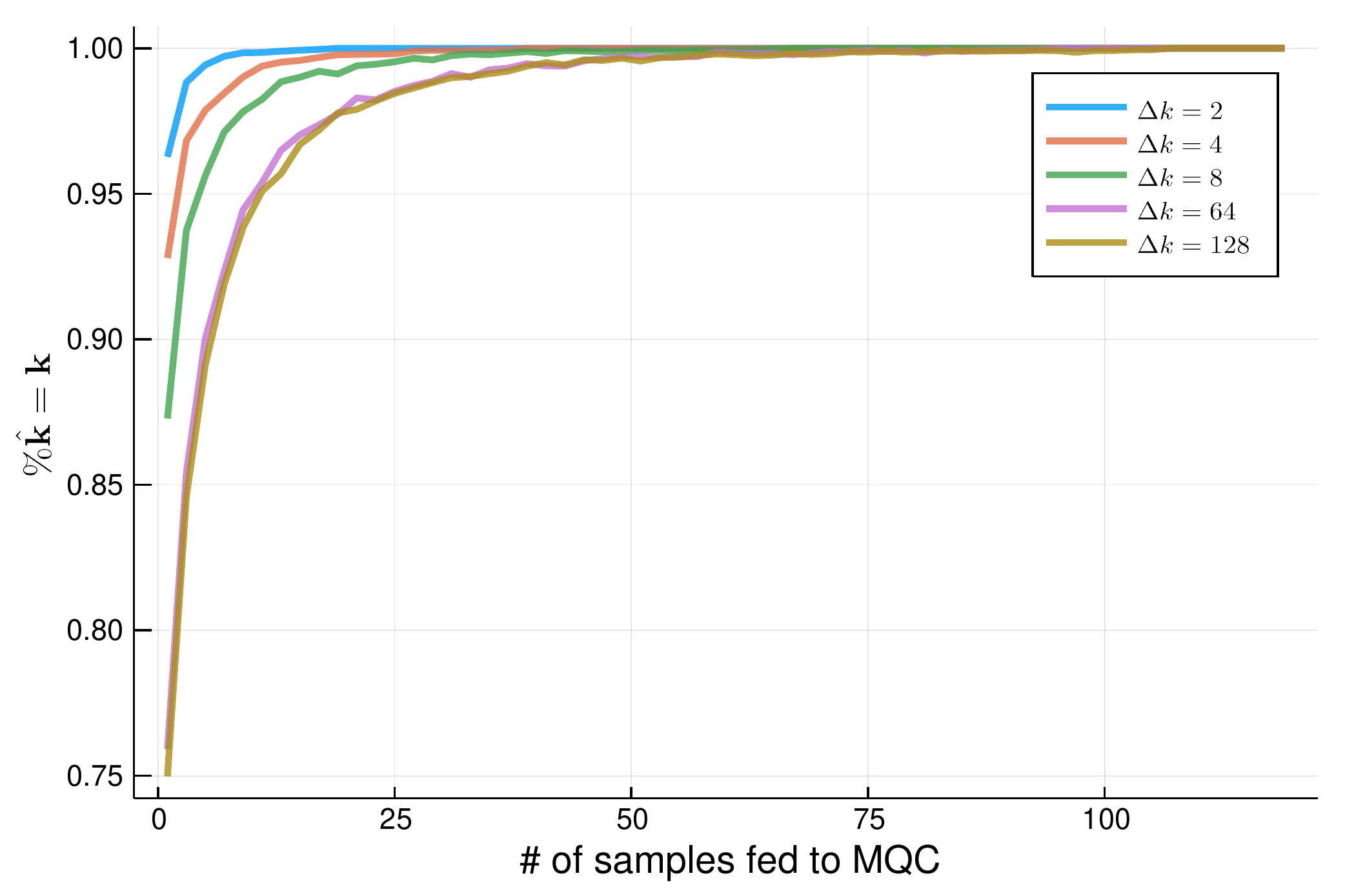}
    \caption{Efficacy of the MQC algorithm for the 1D hydrological inverse problem with $\mathbf{k}$ of length 2030. The accuracy of the MQC algorithm depends on the number of samples being fed into the algorithm as well as the proximity of those samples to the correct answer. As $\Delta k$ increases the number of inputs necessary to achieve a correct result with high accuracy also increases.}
    \label{fig:1d-MQC}
\end{figure}

The 1D problem is non-trivial for MQC, however, and so we use this simple case as a warm-up before discussing the 2D problem.
In Figure~\ref{fig:1d-MQC}, we can see that MQC performance increases dramatically as the number of samples increases to $\approx 25$, but then has a very slow rise beyond that. 
Generating more samples from the quantum annealer hardware and processing them through MQC both operate in polynomial time. However, it is clear from Figure~\ref{fig:1d-MQC} that the resulting increases in accuracy show diminishing returns as the number of samples increases.

Note that MQC performs very similarly for $\Delta k = 64$ and $\Delta k = 128$.
This is because the accuracy of the quantum annealer is effectively the same for all $\Delta k \gtrsim 50$ (as discussed in Section ~\ref{sec:why-large-k-hard}).
Since the MQC algorithm only depends on the number of samples and their accuracy, the MQC effectiveness should also be similar for values of $\Delta k \gtrsim 50$, which we see in Figure~\ref{fig:1d-MQC}.

In conclusion, the 1D version of the hydrological inverse problem is relatively straightforward for the FV and PO algorithms.
MQC requires a modest $\mathcal{O}(100)$ samples to find the optimal solution for any $\Delta k$.

\subsection{Scaling in $\Delta k$}\label{sec:results-2d} 

\begin{figure}[h!]
    \includegraphics[width=\textwidth]{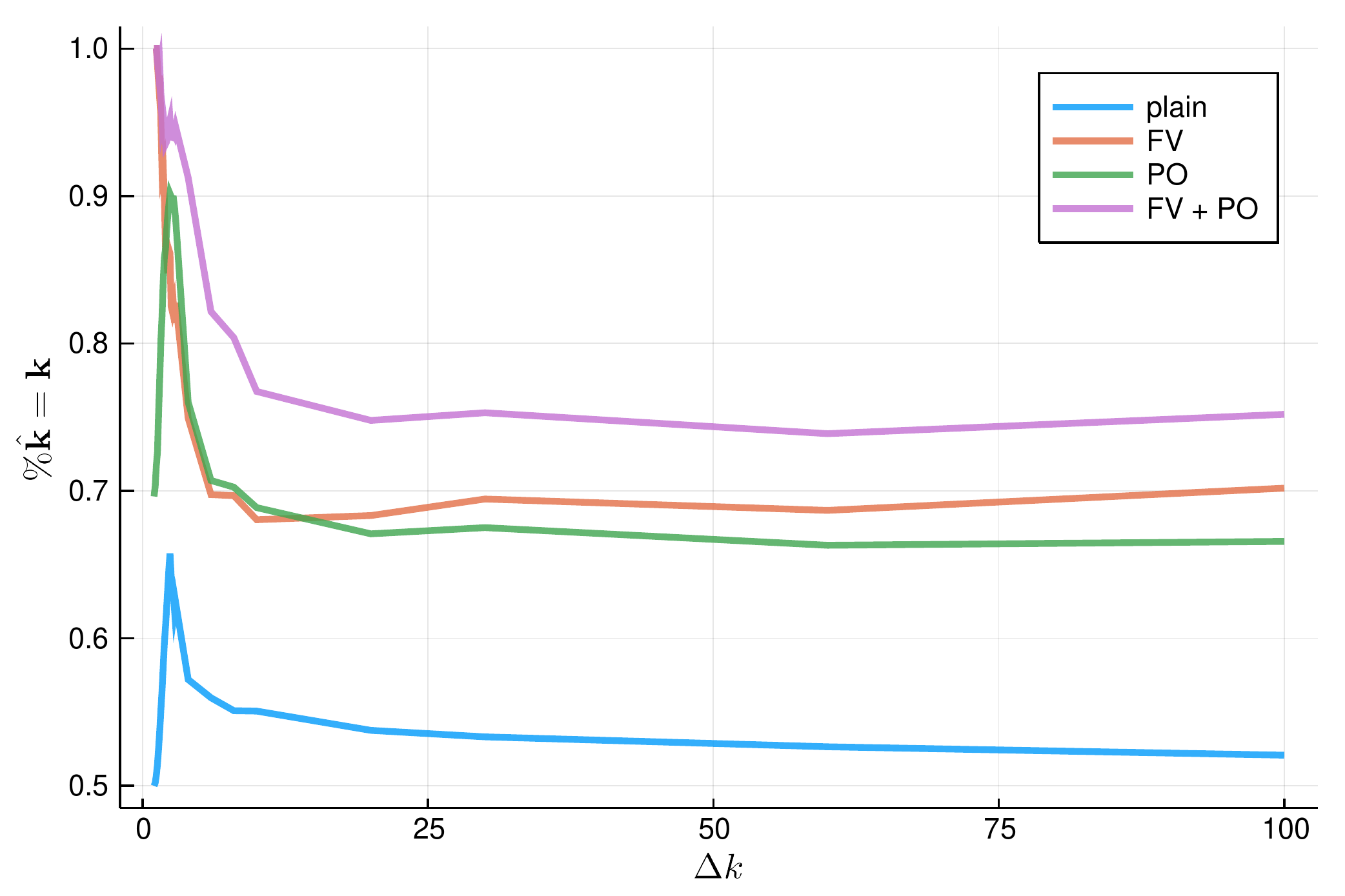}
    \caption{Percentage of $\mathbf{k}$ fixed by FV and/or PO as a function of $\Delta k$ for the 2D hydrological inverse problem. It is unclear why the quantum annealer is best at solving problems in the $\Delta k \approx 2$ regime.
}
    \label{fig:2d-FV+optimize-multi-kH}
\end{figure}

The 2D formulation of the hydrological inverse problem provides more complexity to study the relative efficacy of all three algorithms. 
To start, Figure~\ref{fig:2d-FV+optimize-multi-kH} shows that both FV and PO offer significant improvements over the plain quantum annealer result, with PO slightly outperforming FV in the large $\Delta k$ regime. 
As with the 1D case, we see a plateau in performance for large $\Delta k$, however the plateau value is noticeably lower (compare to Figure \ref{fig:1d-delta-k}) due to the increased difficulty of the problem.

\begin{figure}[h!]
    \includegraphics[width=\textwidth]{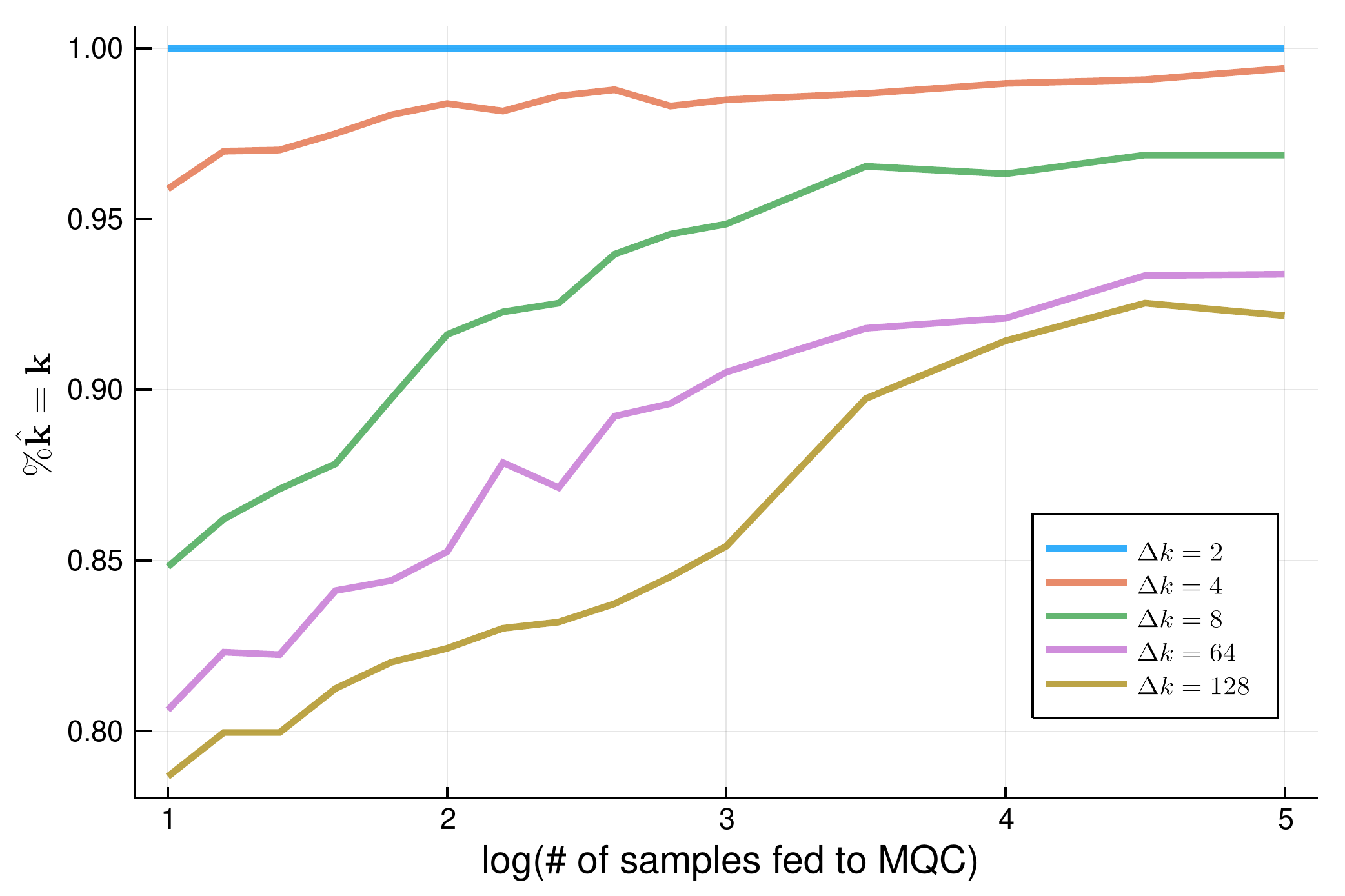}
    \caption{Percentage of $\mathbf{k}$ fixed by MQC as a function of number of FV+PO input samples for the 2D inverse problem.}
    \label{fig:2d-MQC}
\end{figure}

To further illuminate the difficulties underlying the 2D inverse problem, let us briefly consider a custom algorithm.
First, we note that the origin of the poor performance in the 2D case is functionally the same as that of the 1D case discussed in Section~\ref{sec:why-large-k-hard}. 
Namely, some coefficients in the QUBO become orders of magnitude larger than others as $\Delta k$ increases.
One idea for resolving this issue is to solve the QUBO in chunks: have the quantum annealer find a solution for all of the $k_i$, but then ignore the results for the $k_i$ associated with small coefficients as these are effectively noise. 
Then, take the ``trusted'' $k_i$ values (associated with large coefficients) and filter them into the original QUBO.
This generates a new QUBO with only small coefficients, which can then be rescaled and solved again.
Combining the results from the two solutions might then provide a robust answer. This is similar in concept to the algorithm proposed in \cite{karimi2017boosting}.

\begin{figure}[h!]
    \includegraphics[width=\textwidth]{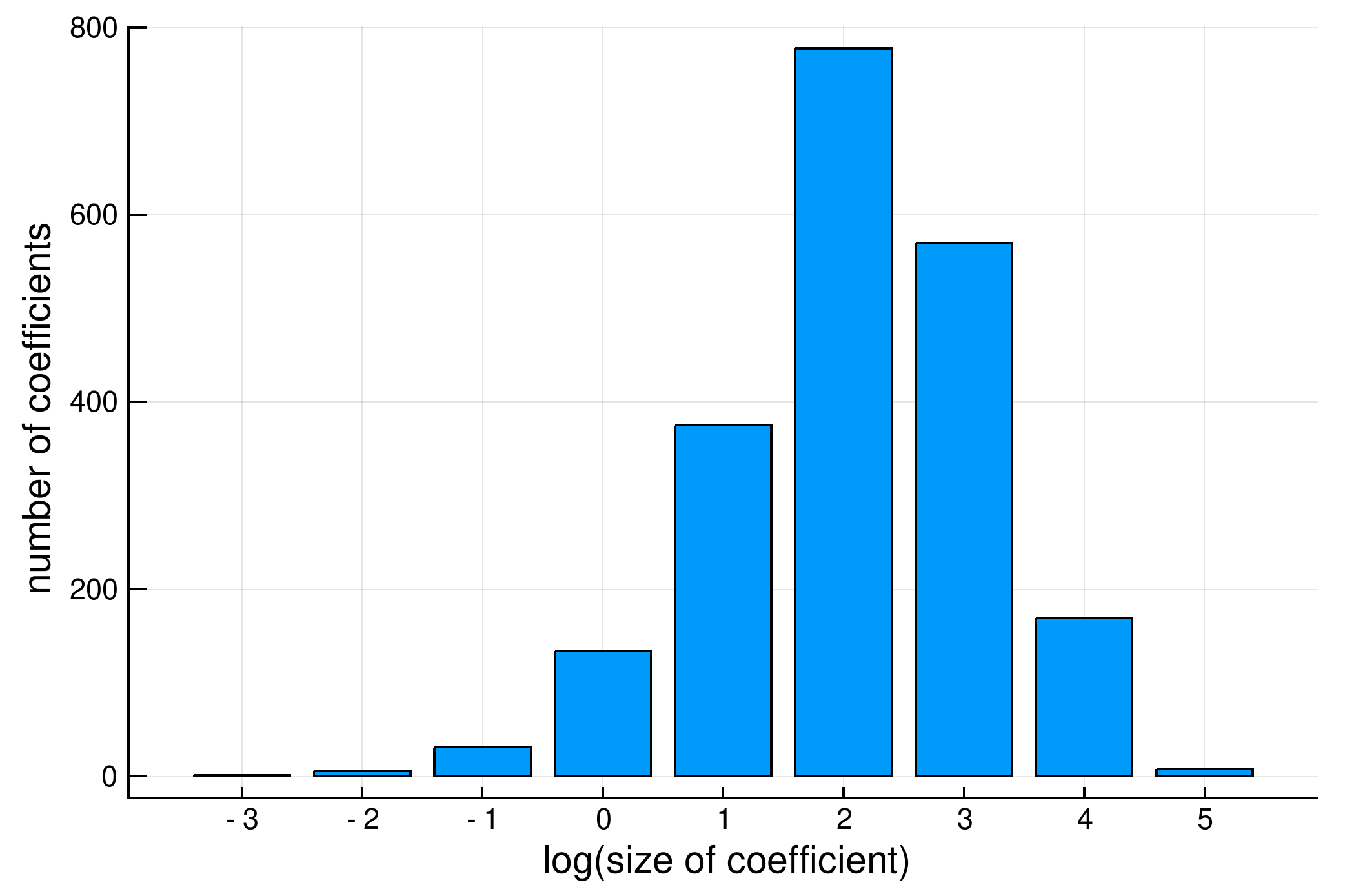}
    \caption{The size of coefficients in the QUBO for a 2D problem with $\Delta k = 50$.}
    \label{fig:2d-coeffs}
\end{figure}

This approach is, in fact, useful in the 1D case, as there are a relatively small number of possible values each coefficient can take, and thus a clear choice of cutoff between ``large'' and ``small'' coefficients -- essentially forming two independent QUBOs.  
However, in the 2D case (whose QUBO coefficients are considerably more complicated, see \cite{omalley2018hydro}) there is a much larger spectrum of possible values. 
Indeed, the QUBO coefficient values vary smoothly over several orders of magnitude, for example see the spectrum in Figure~\ref{fig:2d-coeffs}.

We have heuristically found that there is no good cutoff value.
Any choice leads to a bifurcated pair of QUBOs whose solutions are no longer equivalent to the solution of the original QUBO.

\subsection{Scaling in the size of the grid}\label{sec:scaling}

In this section, we experimentally determine the scaling behavior of the quantum annealer and pre- and post-processing algorithms introduced above for the 2D hydrological inverse problem as the problem size increases.

\begin{figure}[h!]
    \includegraphics[width=\textwidth]{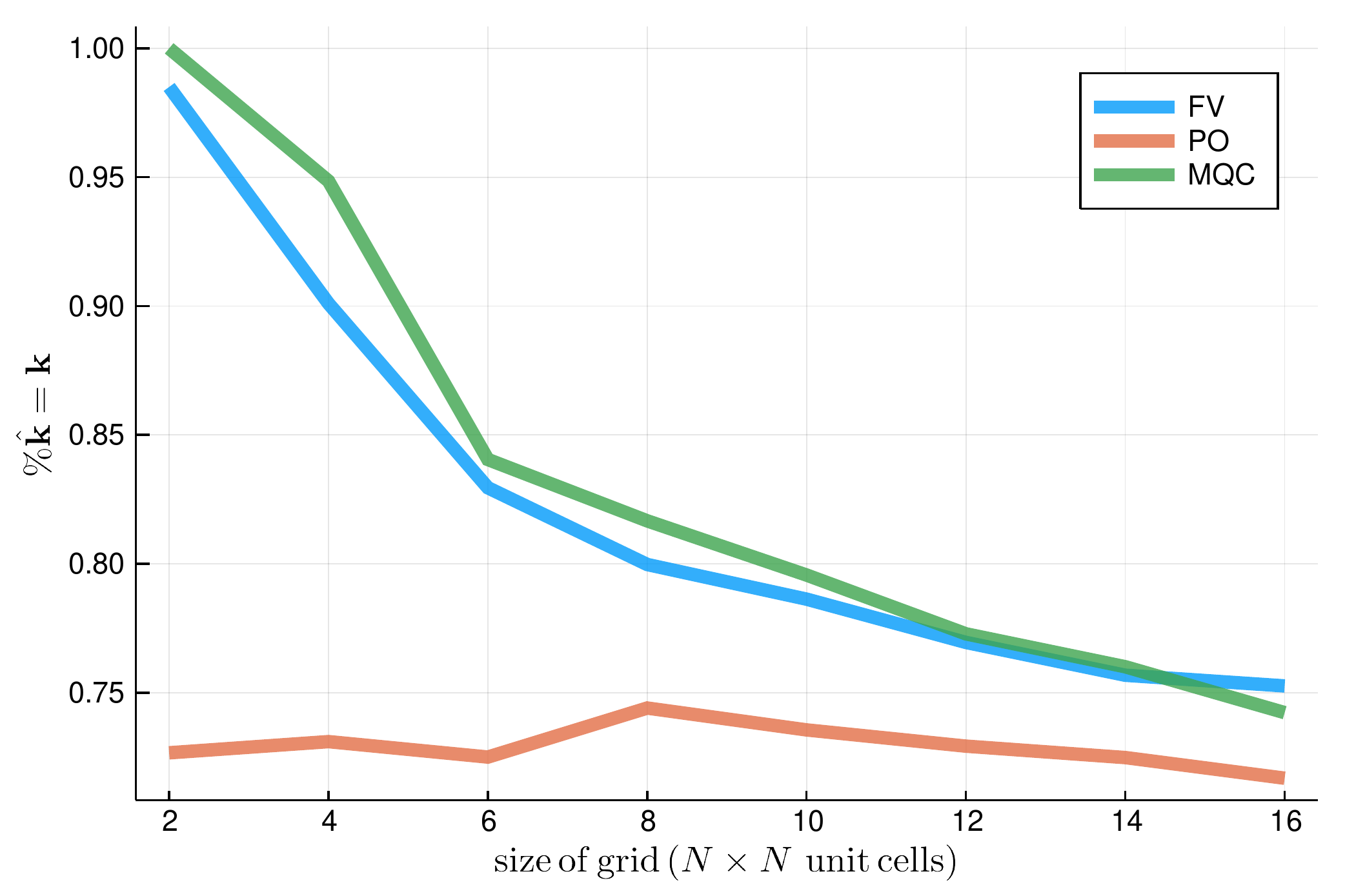}
    \caption{Percentage of $\mathbf{k}$ fixed by FV, PO, and MQC as a function of grid size for $\Delta k = 100$. Accuracy values averaged over 50 different random realizations of an $N \times N$ grid. Each algorithm utilizes 1000 samples from the D-Wave hardware. For FV and PO, we take the best sample; for MQC, we combine the samples. The dip in efficacy for the PO algorithm at small grid size is likely an artifact of boundary conditions for the embedding process.}
    \label{fig:scaling-accuracy}
\end{figure}

Figure \ref{fig:scaling-accuracy} shows the performance of the three algorithms for $\Delta k= 100$ as a function of the $N\times N$ grid.
As discussed previously, we use $N$ to refer to the number of Chimera unit cells that are used, with each unit cell encoding four permeabilities $k^x_{i,j},~k^y_{i,j},~k^x_{i + 1,j},~k^y_{i,j + 1}$ and each permeability included in two unit cells by inter-cell couplers.
FV and MQC perform noticeably better than PO in the small grid regime. However, all approaches appear to asymptote to slightly below 75\% accuracy as the grid size grows. 

\begin{figure}[h!]
    \includegraphics[width=\textwidth]{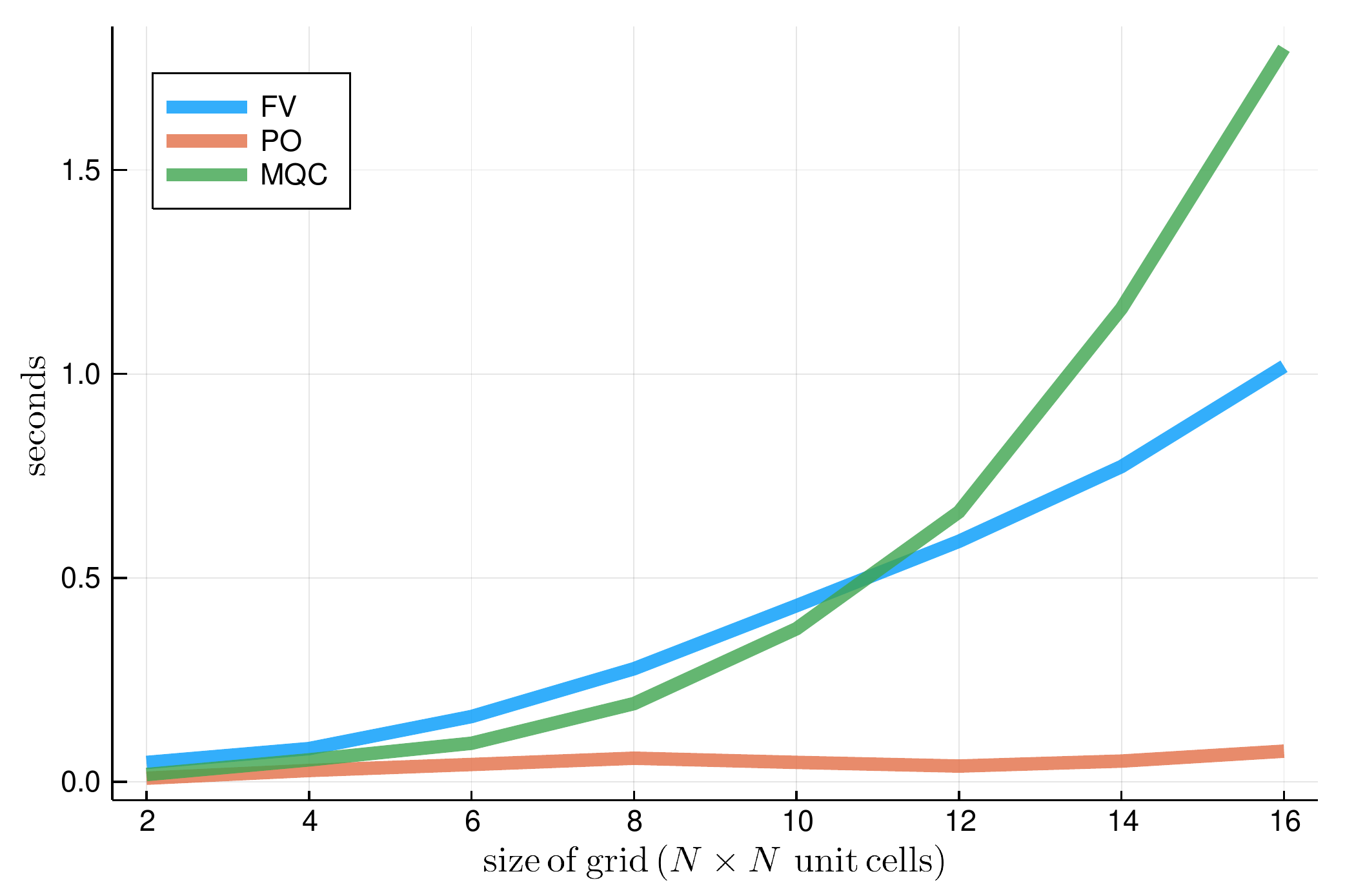}
    \caption{Processing time for FV, PO, and MQC as a function of grid size for $\Delta k = 100$. Time values averaged over the same random grids as in Figure~\ref{fig:scaling-accuracy}.}
    \label{fig:scaling-time}
\end{figure}

Next, we study the computational time for each of the algorithms, see Figure \ref{fig:scaling-time}. PO shows minimal increased time, while MQC and FV become comparatively more expensive as the problem size increases.
MQC is the most expensive for the largest problems considered here, despite having better scaling behavior than FV for very large problems.
We emphasize that all these problems have polynomial time complexity, whereas QUBOs are generally NP-hard.
Therefore, in the scaling sense, these pre- and post-processing algorithms add a modest computational cost to the problem.

\pagebreak

\section{Effects of observational noise}\label{sec:noise}
In this section, we discuss how noisy hydrologic observations affect the QUBO, and by extension the ability of the quantum annealer to correctly determine $\mathbf{k}$.
Observational noise enters the QUBO through the measurements $h_i$. 
We add noise to our simulations by sampling from a distribution $\mathcal{N}(0,\sigma)$ and adding that to each of the $h_i$.
From eqs.~(\ref{eq:qubo-coeffs-a}) and (\ref{eq:qubo-coeffs-b}) it is clear that individual $h_i$ do not affect the QUBO, instead only the differences, $\Delta h_i$, are relevant. 
Furthermore, from eq.~(\ref{eq:delta-h-defs}) it is clear that in the case we are considering ($k_L = 1, ~n_1\approx n_2 \approx n/2$), $\Delta h_i \in [-2,0]$. 
This then means that the scale of the ``signal,'' i.e. the difference in observed hydraulic pressure between adjacent measurement points, is $\mathcal{O}(1)$.

\begin{figure}[h!]
    \includegraphics[width=\textwidth]{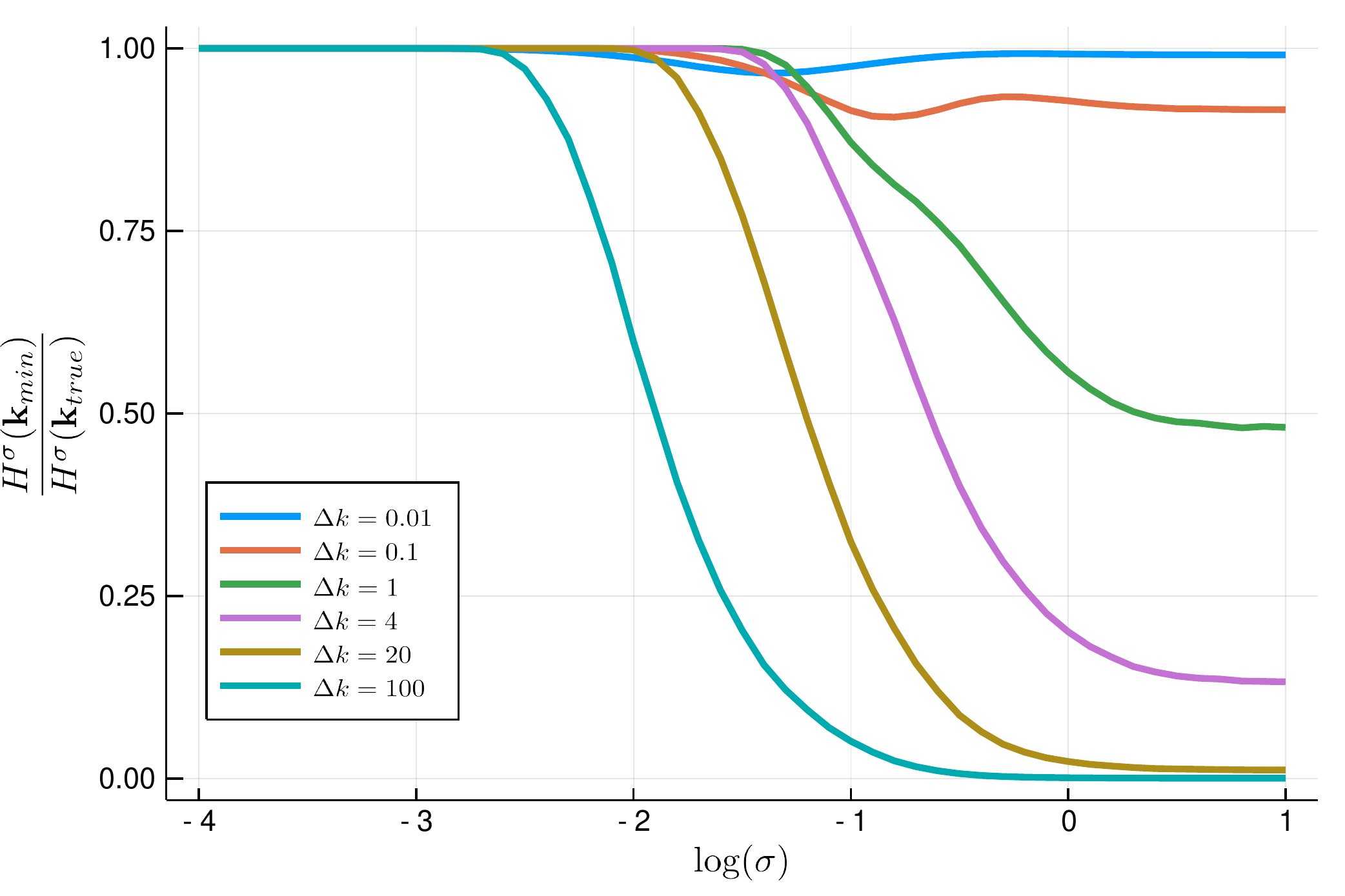}
    \caption{Adding noise to the 1D QUBO produces a new Hamiltonian $H^\sigma$ with a ground state $\mathbf{k}^\sigma_{min}$ that may differ from $\mathbf{k}_{true}$, i.e. $H^\sigma(\mathbf{k}^\sigma_{min}) > H^\sigma(\mathbf{k}_{true})$. This figure shows the ratio between these two energy levels as a function of noise for several values of $\Delta k$.}
    \label{fig:1d-noise}
\end{figure}

How does the amount of noise (characterized by $\sigma$) affect the QUBO?
First, let us describe our notation: recall that we use $\mathbf{k}_{true}$ to denote the collection of synthetic permeabilities describing the ground truth of an aquifer. 
We use this to generate the corresponding noiseless ``observations'' $h_i$, which we then use to construct the QUBO Hamiltonian $H$ (as in eq.~(\ref{eq:dwave-hamiltonian})).
When we add observational noise to the $h_i$, we are then changing $H$ as well, which we then denote by $H^\sigma$.
This new QUBO has a new ground state, which we term $\mathbf{k}^\sigma_{min}$.
For $\sigma = 0$, $\mathbf{k}^0_{min} = \mathbf{k}_{true}$, but in general we will have $H^\sigma(\mathbf{k}^\sigma_{min}) \le H^\sigma(\mathbf{k}_{true})$. 
The larger the difference is between $H^\sigma(\mathbf{k}^\sigma_{min})$ and $H^\sigma(\mathbf{k}_{true})$, the less likely the quantum annealer will produce results that are near $\mathbf{k}_{true}$.
In Figure~\ref{fig:1d-noise} we characterize the degree to which adding noise affects the minimum energy level of the QUBO Hamiltonian for the 1D case.
We have studied the 1D case here because we can use FV to determine the exact ground state, $\mathbf{k}^\sigma_{min}$, thus allowing for direct comparison with the original ground state $\mathbf{k}_{true}$.

When $\Delta k$ is small, all of the energy states of the Hamiltonian are close together, so a completely random $\mathbf{k}$ likely has energy near that of the ground state. 
As $\Delta k$ grows, the energy states become more gapped.
Adding sufficient noise can produce a Hamiltonian whose ground state energy is far less than that of the ideal solution from a hydrologic perspective (i.e., $\mathbf{k}_{true}$).
This further highlights the difficulty of working with a large $\Delta k$.

\begin{figure}[h!]
    \includegraphics[width=\textwidth]{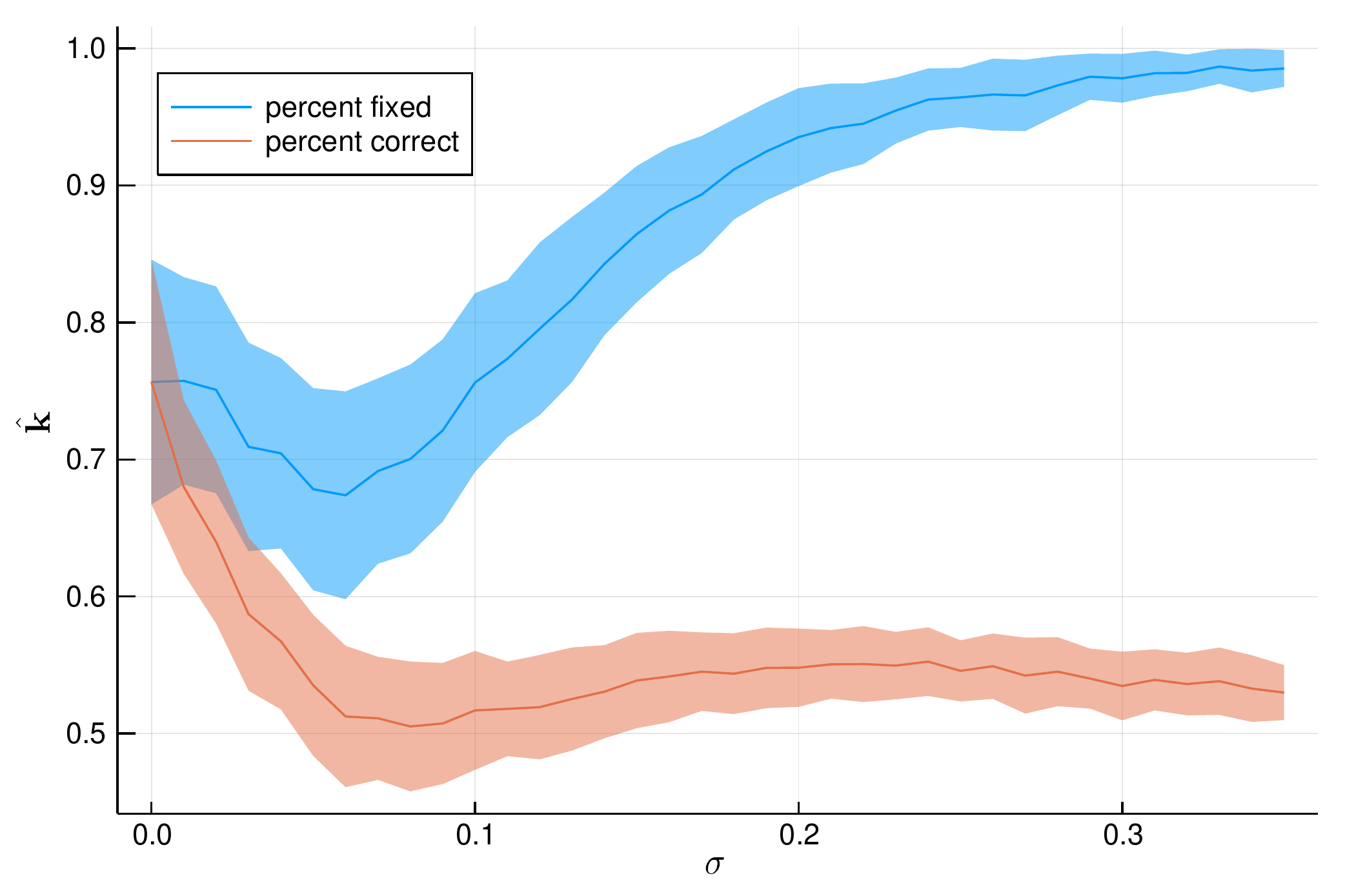}
\caption{Percentage of $\mathbf{k}$ fixed by FV as a function of noise for $\Delta k = 1$ in the 2D hydrological inverse problem. The shaded bar represents one standard deviation of results.}
    \label{fig:2d-FV}
\end{figure}

An interesting feature of the QUBO for the 2D version of eq.~(\ref{eq:dwave-hamiltonian}) is that randomly chosen values for $h_i$ produce a Hamiltonian which is very likely solvable by the FV algorithm, as shown in Figure~\ref{fig:2d-FV}.
In other words, the more noise added to the $h_i$ values, the greater the percentage of $\mathbf{k}$ that FV can determine. 
But, of course, as the noise increases, the solution arrived at by FV further deviates from the desired ground truth. 
It is worth studying more why exactly QUBOs of this form become easier to solve by FV in this very noisy regime. 

\begin{figure}[h!]
    \includegraphics[width=\textwidth]{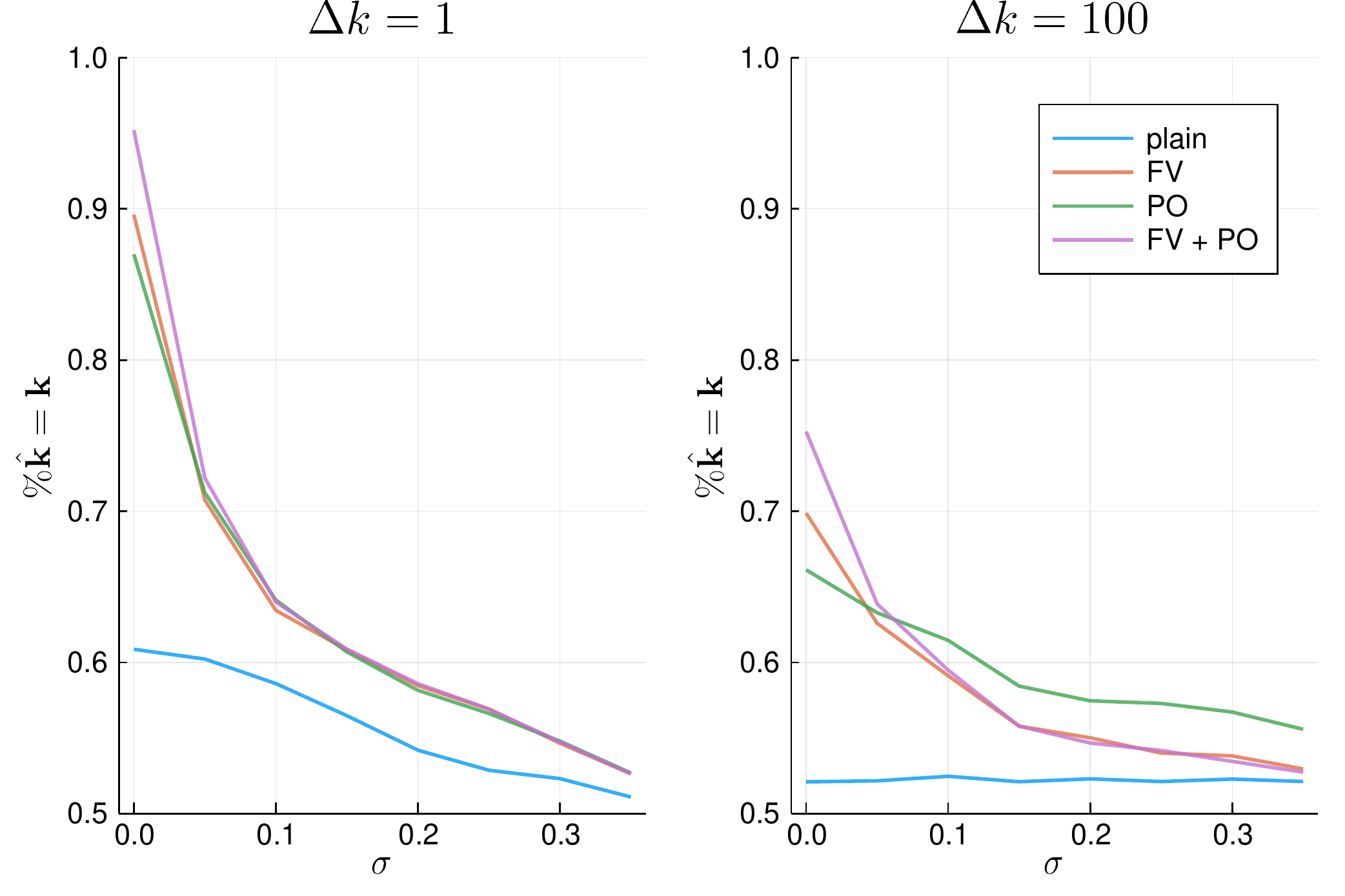}
\caption{Percentage of $\mathbf{k}$ determined by FV and/or PO as a function of noise $\sigma$ for $\Delta k = 1$ and $\Delta k = 100$. PO alone performs best for large $\Delta k$ and $\sigma$.}
    \label{fig:2d-FV+optimize}
\end{figure}

An important consequence of the increased power of FV in the noisy regime is that for large $\Delta k$ and high noise, PO performs better than both approaches involving FV -- see Figure~\ref{fig:2d-FV+optimize}.
FV gets very close to the true minimum of the QUBO, but due to the noise the true minimum ($\mathbf{k}^\sigma_{min}$) is different from the ground truth ($\mathbf{k}_{true}$).
Just using PO allows the quantum annealer to explore more solutions that are sub-optimal for the noisy QUBO but might (by chance) correspond more closely with reality.

\section{Conclusion}\label{sec:conclusion}

In this paper, we have further explored the hydrological inverse problem in the context of quantum annealing hardware, such as the D-Wave 2000Q. 
Specifically, we have focused on the problem of decreased performance by the quantum annealer as the gap between the permeabilities of a porous medium, $\Delta k$, increases. 
We found that performance decreases because of the form of the QUBO Hamiltonian for the hydrologic inverse problem, which contains terms that get very large as $\Delta k$ increases.
These terms dominate lesser terms in the QUBO, and these lesser terms are forced into the noise of the quantum annealer hardware, thus reducing sensitivity to terms that are multiplied by smaller coefficients. 

Several pre- and post-processing techniques were employed to ameliorate this problem. 
Our focus was on general-purpose algorithms to gain insights for broader classes of problems. 
We found that the classical pre-processing technique (here termed FV), involving Roof-duality and strong persistencies, was effective, solving the 1D problem exactly. FV was partially effective at solving the noiseless 2D problem, and interestingly fixed a greater number of variables as more noise was added to the system\footnote{Of course the values determined by FV did not necessarily agree with the actual values of $\mathbf{k}_{true}$ due to the effects of noise}.
The D-Wave API post-processing technique ``optimization'' (here termed PO) was also able to solve the 1D problem exactly. It was also about as effective as FV in the 2D case (and as opposed to FV, it did not determine more variables as noise was added).
Their combined effect was better than each individually, but still failed to exactly solve the 2D problem for reasonably large values of $\Delta k$.
It should be emphasized, however, that getting an exactly correct inverse result is not at all expected in hydrologic inverse problems -- getting an answer that is fairly close to the ground truth is generally considered a success from a hydrologic perspective.

We also studied the post-processing technique of multi-qubit correction, or MQC. 
In 1D, this tool was able to solve the problem with minimal computational overhead, however for 2D the number of samples needed to solve problems exactly with large $\Delta k$ grew prohibitive.
Still, MQC improved the results of the FV and PO samples to above $90\%$ accuracy with relatively few samples.

In addition to analyzing how the accuracy of the algorithms scaled in $\Delta k$, we also studied scaling in the problem size. 
PO exhibited very little variation in accuracy or time as problem size increased, however the accuracy was slightly worse than either FV or MQC. 
FV and MQC were of similar accuracy, with MQC exhibiting the worst scaling of the bunch. 
We conclude that to obtain the best results for this hydrologic inverse problem with a combination of quantum annealing and classical computation is to utilize FV and PO and then run MQC on as many samples as time allows. 

This work raises several questions. 
First, the efficacy of FV in the noisy 2D case is surprising and unexplored. 
Second, we briefly described a na\"ive attempt to design a custom algorithm for improving the quantum annealer results for this specific problem, but custom algorithms that are tailored to this problem merit further consideration.
Third, there are several other options on the 2000Q hardware, such as reverse annealing and anneal offsets, which have the potential to improve upon the results here. 
And finally, it would be interesting to test these methods on upcoming D-Wave hardware with decreased noise at the qubit level.
This decreased noise would allow greater resolution in QUBOs with coefficients of different scales, such as those studied here. 
Our preliminary tests showed an improvement in the performance of the ``plain'' quantum annealer.

\bibliography{refs}{}
\bibliographystyle{unsrt}

\end{document}